\catcode`\@=11

\def\citen#1{\if@filesw \immediate\write \@auxout {\string\citation{#1}}\fi%
\@tempcntb\m@ne \let\@h@ld\relax \def\@citea{}%
\@for \@citeb:=#1\do {\@ifundefined {b@\@citeb}%
    {\@h@ld\@citea\@tempcntb\m@ne{\bf ?}%
    \@warning {Citation `\@citeb ' on page \thepage \space undefined}}%
    {\@tempcnta\@tempcntb \advance\@tempcnta\@ne
    \setbox\z@\hbox\bgroup\ifcat0\csname b@\@citeb \endcsname \relax
    \egroup \@tempcntb\number\csname b@\@citeb \endcsname \relax
    \else \egroup \@tempcntb\m@ne \fi \ifnum\@tempcnta=\@tempcntb
    \ifx\@h@ld\relax \edef \@h@ld{\@citea\csname b@\@citeb\endcsname}%
    \else \edef\@h@ld{\hbox{--}\penalty\@highpenalty
    \csname b@\@citeb\endcsname}\fi
    \else \@h@ld\@citea\csname b@\@citeb \endcsname \let\@h@ld\relax \fi}%
\def\@citea{,\penalty\@highpenalty\hskip.13em plus.13em minus.13em}}\@h@ld}
\def\@citex[#1]#2{\@cite{\citen{#2}}{#1}}%
\def\@cite#1#2{\leavevmode\unskip\ifnum\lastpenalty=\z@\penalty\@highpenalty\fi%
  \ [{\multiply\@highpenalty 3 #1%
  \if@tempswa,\penalty\@highpenalty\ #2\fi}]}   %
\makeatother 

\def\A             {\mbox{$\liefont A$}}
\def\AA            {\mbox{$\liefont A\indA$}}
\def\AE            {\mbox{$\liefont A\indE$}}
\def\ae            {\theta}
\newcommand\Ad[1]  {{\rm Ad}^{}_{#1}}
\def\aff           {affine Lie algebra}

\def\alg           {algebra}
\def\AS            {\mbox{$\liefont A\indS$}}

\def\auto          {automorphism}
\newcommand\Aut[1] {\mathrm{Aut}({#1})}
\def\Autj          {\Aut {\F\indj}}
\def\ax            {\xi}
\def\axt           {\tilde{\xi}}
\def\az            {\zeta}
\def\AZ            {\mbox{$\liefont A\indZ$}}
\renewcommand\b[2] {b^{#1}_{#2}}
\newcommand\B[1]   {B^{}_{#1}}
\newcommand\bc[2]  {a^{#1}_{#2}}
\def\be            {\begin{equation}}
\def\bearl         {\begin{array}{l}}
\def\bearll        {\begin{array}{ll}}
\def\bearlll       {\begin{array}{lll}}
\def\bfe           {{\bf1}}
\newcommand\bogzn[1]{\zeta_{#1}}
\def\bogzz         {\vartheta}
\def\bosn          {\mbox{${\sf Bos}_{N}$}}
\renewcommand\c[2] {c^{#1}_{#2}}
\newcommand\C[1]   {C^{}_{#1}}
\def\car           {{\sf CAR}}
\def\caR           {{\sss({\sf CAR})}}
\def\CAR           {canonical an\-ti-com\-mu\-ta\-tion relation}
\def\cara          {CAR algebra}
\def\cari          {{\sf CAR}(I)}

\newcommand\cd[1]  {{\Delta(#1)}}

\def\cdim          {conformal dimension}
\def\ce            {\mbox{$c\eq1$}}
\def\cE            {{\sf Bos}}
\def\cft           {conformal field theory}
\def\cfts          {conformal field theories}
\def\chiA          {\chii\IndA}
\def\chic          {\chii^{{\sss({\sf Orb})}}}
\newcommand\chid[1]{\chii^{\sss[\dihed]}_{#1}}
\def\chido         {\chii^{\sss[\dihed]}_\dno}
\def\chidc         {\chii^{\sss[\dihed]}_\dnc}
\def\chids         {\chii^{\sss[\dihed]}_\dns}
\def\chidv         {\chii^{\sss[\dihed]}_\dnv}
\def\chii          {\raisebox{.15em}{$\chi$}}
\def\chim          {\chii^{\sss\ZZ}_-}
\newcommand\chio[1]{\chii^{\sss[\Ozw]}_{#1}}
\def\chip          {\chii^{\sss\ZZ}_+}
\def\chiS          {\chii\IndS}
\def\chiz          {\chii^{\sss(\SO)}}

\def\ci            {\cite{bofu}}
\def\Circ          {\,{\circ}\,}

\newcommand\coef[1]{{\rm p}^{}_{#1}}
\def\complex       {{\dl C}}

\def\cwei          {conformal weight}

\def\dcar          {\tild{\cal H}^{\sss(\car)}}
\newcommand\del[2] {\delta_{#1,#2}}
\newcommand\Del[2] {\delta^{#1,#2}}
\def\df            {\,{:=}\,}
\def\dh            {\tild{\cal H}}
\def\dha           {\tild{\cal H}\Inda}
\def\dhA           {\tild{\cal H}\IndA}
\def\dhs           {\tild{\cal H}\Inds}
\def\dhS           {\tild{\cal H}\IndS}

\def\dihed         {{{\dll D}_{2N}}}
\def\Dihed         {\mbox{${\dll D}_{2N}$}}
\def\dihedh        {{\hat{\dll D}_{2N}}}
\def\dll           {\rm }
\def\dn            {\mbox{${\dll Q}_{N}$}}
\def\dne           {\mbox{${\dll Q}_1$}}
\def\Dn            {{{\dll Q}_{N}}}
\def\dnc           {{\rm c}}
\def\dnd           {\mbox{${\dll Q}_{N}^{\sss({\rm diag})}$}}
\def\dnh           {\mbox{$\hat{{\dll Q}}_{N}$}}
\def\Dnh           {{\hat{{\dll Q}}_{N}}}
\def\dno           {\circ}
\def\dns           {{\rm s}}
\def\dnv           {{\rm v}}
\let\dstyle=\displaystyle

\def\ee            {\end{equation}}
\def\eE            {{\rm e}}
\def\eear          {\end{array}}
\def\eeta          {\eta}

\def\emt           {energy-momentum tensor}

\def\eq            {\,{=}\,}
\def\Equiv         {\,{\equiv}\,}
\newcommand\erf[1] {(\ref{#1})}
\newcommand\Erf[2] {(\ref{#1#2})}
\def\F             {\mbox{$\liefont F$}}
\def\Fa            {\mbox{$\liefont F\inda$}}
\def\FA            {\mbox{$\liefont F\indA$}}
\def\FE            {\mbox{$\liefont F\indE$}}
\def\fielda        {field algebra}
\def\findim        {finite-dimensional}
\def\Fj            {\mbox{$\liefont F\indj$}}
\def\fline         {{~}\\[1 mm]\noindent ------------------\\[1 mm]}

\def\fock          {\mbox{$\hcar$}}
\newcommand\Frac[2]{\mbox{\large$\frac{#1}{#2}$}}
\def\Fs            {\mbox{$\liefont F\inds$}}
\def\FS            {\mbox{$\liefont F\indS$}}

\def\futnote#1     {\footnote{~#1}\ }
\def\FZ            {\mbox{$\liefont F\indZ$}}

\def\gdual         {set of irreducible characters}

\def\glnh          {\mbox{$\widehat{\liefont{gl}}(2N)$}}

\def\H             {\mbox{$\cal H$}}

\def\hA            {{\cal H}\IndA}
\def\Ha            {{\cal H}\Inda}

\def\half          {\mbox{\large$\frac12$}\,}

\def\hcar          {{\cal H}^{\sss(\car)}}

\def\hcarm         {{\cal H}^{\sss(\car)}_-}
\def\hcarn         {{\cal H}^{\sss(\car,2N)}}
\def\hcarp         {{\cal H}^{\sss(\car)}_+}
\def\he            {{H\indE}}
\def\hE            {{\cal H}\IndE}
\def\HE            {\mbox{${\cal H}\IndE$}}

\def\Hm            {{\cal H}}

\def\hS            {{\cal H}\IndS}
\def\Hs            {{\cal H}\Inds}

\newcommand\hsp[1] {\mbox{\hspace{#1em}}}
\def\hw            {highest weight}

\def\hy            {$\mbox{-\hspace{-.66 mm}-}$}
\def\hz            {{H\indZ}}
\def\hZ            {{\cal H}\IndZ}
\def\HZ            {\mbox{${\cal H}\IndZ$}}

\def\id            {\mbox{\sl id}}

\def\ii            {{\rm i}}
\def\ihwm          {irreducible highest weight module}
\def\iN            {\!\in\!}
\newcommand\ind[3] {\mathrm{ind}^{#1}_{#2}(#3)}
\def\inda          {_{{\sss(\cE)}}}
\def\indA          {_{{\sss(\cE;2N)}}}
\def\Inda          {^{{\sss(\cE)}}}
\def\IndA          {^{{\sss(\cE;2N)}}}
\def\inde          {_{{\sss(1)}}}
\def\indE          {_{{\sss(1)}}}
\def\IndE          {^{{\sss(1)}}}
\def\indj          {_{{\sss(j)}}}
\def\Indj          {^{{\sss(j)}}}
\def\inds          {_{{\sss(\SU)}}}
\def\indS          {_{{\sss(\suN)}}}
\def\Inds          {^{{\sss(\SU)}}}
\def\IndS          {^{{\sss(\suN)}}}
\def\indz          {_{{\sss(2)}}}
\def\indZ          {_{{\sss(2)}}}
\def\IndZ          {^{{\sss(2)}}}

\def\irmod         {irreducible module}
\def\irrep         {irreducible representation}
\newcommand\J[3]   {J^{#1#2}_{#3}}
\newcommand\je[1]  {J^{}_{#1}}
\def\Je            {J}
\newcommand\jet[1] {\tilde J^{}_{#1}}
\def\Jet           {\tilde J}
\newcommand\jo[3]  {I^{#1#2}_{#3}}
\def\Js            {\mathcal{J}^*}

\def\km            {Kac\hy Moo\-dy }
\def\kma           {Kac\hy Moo\-dy algebra}

\long\def\labl#1   {\label{#1}\ee \ifnum\draftcontrol=1
                   \mbox{ }\\[-12 mm]\query{#1}\\[5 mm] \fi}
\long\def\Labl#1#2 {\label{#1#2}\ee\ifnum\draftcontrol=1
                   \mbox{ }\\[-12 mm]\query{#1#2}\\[5 mm] \fi}
\newcommand\Lb[1]  {L_{#1}\Inda}
\newcommand\LB[1]  {L_{#1}\IndA}
\def\Ldots         {,...\,,}
\def\lE            {\,{\le}\,}

\def\li            {\mbox{$\Lambda_{(a)}$}}
\def\lI            {{\Lambda_{(a)}}}
\def\LIC           {L^2(I;\!\complex^{2N})}
\def\lie           {Lie algebra}

\newcommand\Lj[1]  {\Lambda_{(#1)}}
\newcommand\Ll[1]  {L_{\Lambda_{(#1)}}\Inds}

\def\lLb           {\mbox{\large(}}

\def\LLb           {\mbox{\Large[}}
\def\Lll           {L_{\Lambda}\Inds}
\def\LLl           {L_{\Lambda}\IndS}
\def\lo            {\Lambda_{(0)}}

\def\lRb           {\mbox{\large)}}
\def\LRb           {\mbox{\Large]}}

\def\LSC           {L^2(S^1;\!\complex^{2N})}
\def\lT            {\,{<}\,}
\newcommand\Lu[1]  {L_{#1}\oue}

\def\mi            {\,{-}\,}

\newcommand\N[3]   {N_{\!#1#2}^{\,\ #3}}
\def\natnum        {{\dl N}}
\def\natnumo       {{\dl N}_0}
\def\Ne            {\,{\not=}\,}

\newcommand\normord[1] {\,\raisebox{.033em}{\large\bf:}#1
                   \raisebox{.033em}{\large\bf:}\,}

\def\obsa          {observable algebra}
\def\od            {^\Dn_{}}
\def\odd           {^{\Dn{\sss({\rm diag})}}_{}}
\def\odihed        {^\dihed_{}}
\def\odtd          {^{\Dn\times\Dn}_{}}
\def\og            {^G_{}}
\def\oh            {^H_{}}

\def\ohe           {^\he_{}}

\def\ohz           {^\hz_{}}
\def\ok            {^{K_N}_{}}

\def\onedim        {one-dimensional}
\def\onehalf       {\mbox{$\frac12$}}
\def\oneton        {1,2,...\,,2N}
\def\onetond       {1,2,...\,,N{-}1}
\def\onetone       {1,2,...\,,2N{-}1}

\newcommand\Oplusd[1]{\!\displaystyle\bigoplus_{#1\in\Dnh}\!}
\newcommand\oplusg[1]{\displaystyle\bigoplus_{#1\in\hat G}}
\newcommand\oplusgt[1]{\bigoplus_{#1\in\hat G}}
\newcommand\oplush[1]{\displaystyle\bigoplus_{#1\in\hat H}}
\newcommand\oplusz[1]{\displaystyle\bigoplus_{#1\in\zet_{2N}}}
\newcommand\oplusZ[1]{\displaystyle\bigoplus_{#1\in\zet}}
\newcommand\Oplusz[1]{\bigoplus_{#1\in\zet_{2N}}}
\def\orbn          {\mbox{${\sf Orb}_N$}}
\def\otim          {\raisebox{.07em}{$\scriptstyle\otimes$}}
\def\otiM          {\,{\otim}\,}
\def\Otimes        {\,{\otimes}\,}
\def\OTimes        {{\otimes}}

\def\oUe           {^{\sss[{\rm U}(1)]}}
\def\oue           {^{\sss({\liefont u}(1))}}
\def\oz            {^\Zn_{}}
\def\ozd           {^{\Zn{\sss({\rm diag})}}_{}}
\def\oztz          {^{\Zn\Times\Zn}_{}}
\def\Ozw           {{\mathrm{O}(2)}}
\def\Ozwh          {\widehat{\mathrm{O}(2)}}
\def\ozz           {^\Zz_{}}
\def\P             {{\phantom|}}
\def\pid           {\pi^{\sss(\Dn)}}
\def\pidc          {\pi^{\sss(\Dn)}_\dnc}
\def\pido          {\pi^{\sss(\Dn)}_\dno}
\def\pids          {\pi^{\sss(\Dn)}_\dns}
\def\pidv          {\pi^{\sss(\Dn)}_\dnv}
\def\pidi          {\pi^{\sss(\dihed)}}
\def\pidic         {\pi^{\sss(\dihed)}_\dnc}
\def\pidio         {\pi^{\sss(\dihed)}_\dno}
\def\pidis         {\pi^{\sss(\dihed)}_\dns}
\def\pidiv         {\pi^{\sss(\dihed)}_\dnv}
\def\pio           {\pi^{\sss(\Ozw)}}

\def\pit           {\tilde\pi}
\def\pizzm         {\pi^{\sss(\Zz)}_-}
\def\pizzp         {\pi^{\sss(\Zz)}_+}
\def\pl            {\,{+}\,}

\def\psI           {\phi}
\newcommand\psif[2]{\psi^{}_{#2}(q)}
\newcommand\Psif[2]{\psi^{}_{#2}}

\def\qft           {quantum field theory}
\def\qfts          {quantum field theories}
\long\def\query#1{\hskip 0pt{\vadjust{\everypar={}\small\vtop to 0pt{\hbox{}%
     \vskip -13pt\rlap{\hbox to 47pc{\hfil{\vtop{\hsize=8pc\tolerance=6000%
     \hfuzz=.5pc\rightskip=0pt plus 3em\noindent#1}}}}\vss}}}}%

\def\reals         {{\dl R}}
\def\rep           {representation}
\def\Rep           {Representation}
\newcommand\res[3] {\mathrm{res}^{#1}_{#2}(#3)}
\def\resp          {respectively}

\def\rmc           {{\rm c}}
\def\rms           {{\rm s}}
\def\rmo           {\circ}
\def\rmv           {{\rm v}}
\def\role          {r\^ole}
\def\rring         {\rep\ ring}
\def\scs           {\scriptstyle}
\newcommand\Sect[2] {\sect{#1}\label{s.#2}
                   \ifnum\draftcontrol=1 \query{s.#2} \fi}

\def\semitimesr    {\begin{picture}(8,8)\put(7.7,0.12){\line(0,1)
                   {5.96}}\put(0,0){$\times$} \end{picture}\, }

\def\Sim           {\,{\sim}\,}

\newcommand\sN[1]  {\Theta^{}_{#1}(q)}
\newcommand\SN[1]  {\Theta^{}_{#1}}
\def\SO            {{\sf SO}}

\def\sonh          {\mbox{$\widehat{\liefont{so}}(2N)$}}
\def\sonhp         {\mbox{$\widehat{\liefont{so}}(2N{+}1)$}}

\newcommand\srf[1] {section \ref{s.#1}}
\def\sss           {\scriptscriptstyle}
\def\SU            {{\sf SL}}
\def\SUEE          {{\rm SU}(1,\!1)}
\def\subseT        {\,{\subset}\,}
\def\sumi          {\sum_{i=1}^{2N}}
\def\sumI          {\sum_{\,i=1}^{2N}}
\newcommand\sumn[1]{\sum_{#1=1}^{2N}}
\newcommand\sumN[1]{\sum_{#1}^{2N}}

\newcommand\summN[1]{{\displaystyle\sum_{\scriptstyle m_1,m_2,...,m_{2N}\in\zet
                   \atop \scriptstyle m_1+m_2+...+m_{2N}=#1}\!\!\!}}
\newcommand\sumz[1]{\displaystyle\sum_{#1\in\zet}}
\newcommand\sumzh[1]{\displaystyle\sum_{#1\in\zet+1/2}\hsp{-.5}}
\newcommand\sumzH[2]{\displaystyle\sum_{\scs#2\atop\hsp{#1}\scs\in\zet+1/2}
                   \hsp{-1.1}}
\newcommand\sumZH[2]{\displaystyle\sum_{\scs#2\atop\hsp{#1}\scs\in\zet+1/2}
                   \hsp{-.7}}
\newcommand\Sumzh[1]{\sum_{#1\in\zet+1/2}}
\def\sun           {\mbox{$\liefont{sl}(2N)$}}
\def\suN           {{{\sf SL};2N}}
\def\sunh          {\mbox{$\widehat{\liefont{sl}}(2N)$}}
\def\suse          {superselection sector}

\def\suzh          {\mbox{$\widehat{\liefont{sl}}(2)$}}
\def\tild          {}
\def\Times         {\,{\times}\,}
\def\twodim        {two-di\-men\-si\-o\-nal}
\renewcommand\u[2] {u^{#1}_{#2}}
\def\Ua            {U\Inda}
\def\UA            {U\IndA}

\def\Ue            {\mbox{${\rm U}(1)$}}
\def\UE            {U\IndE}
\def\ueh           {\mbox{$\widehat{\liefont u}(1)$}}
\def\Uj            {U\Indj}
\def\Us            {U\Inds}
\def\US            {U\IndS}
\def\UZ            {U\IndZ}
\renewcommand\v[2] {v^{#1}_{#2}}
\def\vac           {|\Omega\rangle}
\newcommand\version[1] {\ifnum\draftcontrol=1 \typeout{}\typeout{#1}\typeout{}
                   \vskip3mm \centerline{\fbox{{\tt DRAFT -- #1 -- }
                   {\small\draftdate}}} \vskip3mm \fi}
\def\vi            {\varphi}

\def\VV            {R}
\def\wrt           {with respect to }
\def\wrtt          {with respect to the }

\def\wzwo          {\mbox{${\sf SO}_{2N}$}}
\def\wzwt          {WZW theory}
\def\wzwts         {WZW theories}
\def\wzwu          {\mbox{${\sf SL}_{2N}$}}
\def\x             {{\tilde x}}
\def\y             {{\tilde y}}
\def\zet           {{\dl Z}}
\def\zeth          {\zet{+}1/2}

\def\zn            {\mbox{$\zet_{2N}$}}
\def\Zn            {{\zet_{2N}}}
\def\znd           {\mbox{$\zet_{2N}^{\sss({\rm diag})}$}}
\def\zz            {\mbox{$\zet_2$}}
\def\Zz            {{\zet_2}}
\def\ZZ            {[\zet_2]}

\def\draftdate{\number\month/\number\day/\number\year\ \ \ \hourmin }

\global\def\draftcontrol{0}
\catcode`\@=12

\documentclass[12pt]{article}\usepackage{amssymb,amsfonts}

\newcommand\sect[1] {\section{#1}\setcounter{equation}{0}}

\begin{document}

\let\dl=\bf      \let\liefont=\bf
\let\dl=\mathbb  \let\liefont=\mathfrak


\begin{flushright}  {~} \\[-15 mm]  {\sf hep-th/9705039} \\[1mm]
{\sf DESY 97-077} \\[1 mm]{\sf May 1997} \end{flushright} \vskip 2mm \mbox{~}

\begin{center} \vskip 13mm
{\Large\bf CFT FUSION RULES, DHR GAUGE GROUPS,}\\[2.2mm]
{\Large\bf AND CAR ALGEBRAS}\\[12mm]
{\large Jens B\"ockenhauer} \\[3mm]
University of Wales Swansea, Department of Mathematics\\[.6mm]
Singleton Park, \ Swansea~~SA2 8PP \\[9mm]
{\large J\"urgen Fuchs} $^{\sf X}$ \\[3mm]
DESY\\[.6mm] Notkestra\ss e 85, \ D -- 22603~~Hamburg
\end{center}
\vskip 20mm

\begin{quote}{\bf Abstract}.\\
It is demonstrated that several series of
\cfts, while satisfying braid group statistics, can still
be described in the conventional setting of the DHR theory, i.e.\ their
superselection structure can be understood in terms of a compact DHR gauge
group. Besides theories with only simple sectors, these include
(the untwisted part of) $c\,{=}\,1$ orbifold theories and level two \sonh\
WZW theories. We also analyze the relation between these models and theories of
complex free fermions.
\end{quote}
\vfill {}\fline{} {\small $^{\sf X}$~~Heisenberg fellow}
\newpage


\sect{Introduction}

In local relativistic quantum field theory, the {\em fusion rules\/} encode the
basis independent features of the composition of \suse s.
When the theory enjoys permutation group statistics, as is e.g.\ the case in
four space-time dimensions, then -- under
standard assumptions which are motivated by physical principles such as
causality (see e.g.\ \cite{HAag}) -- the fusion rules can be studied with the
help of the DHR theory \cite{dohr,dohr2,doro0}, implying that the composition
of sectors is governed by a compact group \cite{doro4}, the {\em DHR gauge
group}. More precisely, the fusion ring of the theory is isomorphic to the
\rring\ of the gauge group. On the other hand, \qfts\ in space-times of low
dimensionality generically possess braid group statistics. As a
consequence the \role\ of the gauge group is taken over by a much more
complicated structure, which is commonly called a quantum symmetry, and for
which no generally accepted description is available yet.

The purpose of this paper is to demonstrate that several classes of
low-dimensional \qfts, while satisfying braid group statistics, can
nevertheless be described in the conventional
DHR setting. The models in question are certain rational
\cfts. A necessary prerequisite for such a description to work
is that the statistical dimensions $d_a$ of all sectors are integers,
  \be  d_a \in\zet \quad{\rm for\ all}\ a  \,.  \labl0
In the particular case that all sectors are simple in the sense that they
have statistical dimension 1, the relevant braid group \rep\ is \onedim\ and
the fusion ring is nothing but the group ring of
a finite abelian group. For these `abelian' theories it is rather
straightforward to interpret the relevant abelian group as a DHR gauge group.

However, as we will show in this paper, a DHR interpretation is even possible
for series of rational \qfts\ which satisfy \erf0, but for which sectors with
statistical dimension larger than 1 appear. More specifically, we consider\,%
\futnote{Another example has been studied in the appendix of \cite{rehr5}.}
series of models which are labelled by a natural number $N\iN\natnum$, such that
there are four sectors of statistical dimension 1 and
$N\mi1$ sectors of statistical dimension 2. In \cft, two different
realizations of this series are known: first, the untwisted\,%
\futnote{The term `untwisted' refers to a $\zet_2$-gradation of the full
fusion ring of these \cfts. The $N\mi1$ untwisted sectors form a sub-fusion
ring of the full fusion ring, and it is this subring we consider here.
In addition there are four twisted sectors of statistical dimension $\sqrt N$.
Clearly, with our methods we cannot study these twisted sectors, except
possibly when $N$ is a square (for the latter case see the speculations
at the end of the paper). Note, however, that
in the DHR framework no recourse to concepts like modular invariance
which do not have an immediate physical interpretation is needed.
Accordingly, while the twisted sectors must be included when one wishes
to construct a modular invariant \twodim\ \cft, in our present study of \onedim\
chiral \cfts\ we are free to restrict to the sub-fusion ring of our interest.}
sectors of `orbifold theories' of conformal central charge \ce\
(see e.g.\ \cite{dvvv}), and second, the untwisted sectors of \wzwts\
\cite{knza} that are based on the \sonh\ current \alg\ at level 2, which have
$c\eq2N{-}1$. Both of these realizations can be understood in terms of a 
reduction of other series of \cfts\ which have $2N$ sectors, each of statistical
dimension 1, namely of the \ce\ theories that describe a free boson compactified
on a circle of suitable radius, and of the $c\eq2N{-}1$ \wzwts\ based on
\sunh\ at level 1, \resp. Accordingly, we will start our discussion with an
analysis of those `abelian' theories.

As it turns out, all the theories just mentioned are closely related to
theories of complex\,%
\futnote{There is one other known series of rational fusion rings satisfying
\erf0, namely those describing the untwisted sectors of the level two \sonhp\
\wzwt. Using the results of \ci, these could be analyzed along similar lines,
but involving an odd number of real free fermions.
However, some of the arguments turn out to become more involved, and we
refrain from delving into these complications here.}
free fermions. In establishing these relations,
an important \role\ will be played by various types of DHR gauge groups.
We would like to mention that the description in terms of fermions has
several advantages \wrt a formulation via free bosons, which for these
theories exists as well. For instance,
one only has to deal with polynomials in the basic Fourier
modes rather than with the exponentials that appear in the vertex operators of
the bosonic formulation. In particular, there is a rather
simple characterization of the observables, which will be described in \srf O.

Concerning the free fermion theories, it is worth while to be a bit
more specific already at this introductory stage.
We consider a Fock representation of the \CAR s (CAR) which is characterized
as follows. We fix (once and for all) a positive integer $N$.
On the Fock space $\hcar\equiv\hcarn$ there operate the Fourier modes $\b ir$
and $\c ir$ ($i\iN\{\oneton\}$ and $r\iN\zeth$) of $2N$ complex free fermions.
These modes satisfy the relations
  \be  \bearl  \{\b ir,\c js\} = \del ij\, \del r{-s} \, \bfe\,, \\[.6em]
  \{\b ir,\b js\} = 0 = \{\c ir,\c js\} \,,  \eear \labl C
and there is a $^*$-operation (an involutive automorphism), which acts on the
modes as
  \be  (\b ir)^* = \c i{-r} \,.  \labl*
The Fock space \fock\ itself is defined by the properties that it contains
a unique (up to a phase) vacuum vector $\vac\iN\hcar$ on which the
modes $\b ir$ and $\c ir$ with positive index $r$ act as annihilation operators,
i.e.\ for all $i\eq\oneton$ and all $r\iN\natnumo + 1/2$ we have
$\b ir\vac = \c ir \vac\eq0$, while the modes with negative index $r$ act as
creation operators such that their successive action on the
vacuum provides a dense subspace.

Given the fermion Fourier modes $\b ir$ and $\c ir$, we can define
local and global CAR algebras as follows. For $f\iN\LSC$ we set
  \be  b(f):= \sumi \sumzh r f_{-r}^i \b ir \qquad{\rm and}\qquad
  c(f):= \sumi \sumzh r f_{-r}^i \c ir \,,  \ee
where $f^i(z)\eq\Sumzh r f^i_{-r} z^r$
is the Fourier decomposition of the $i$th component of
the function $f$ (the circle $S^1$ is considered as the unit circle in the
complex plane, so that $z\eq\eE^{\ii\phi}\iN S^1$ with $-\pi \lT \phi \lE \pi$).
For open intervals $I\subseT S^1$ we define the
local \cara s $\cari$ to be the von Neumann algebras generated by elements
$b(f)$ and $c(g)$ with $f,g$ having support in $I$,
  \be \cari:= \left\{ b(f),\,c(g) \,|\, f,g\iN\LIC \right\} ''\,. \ee
(The prime stands for the commutant in the \alg\ of bounded operators on
$\hcar$. Note that the bicommutant of a set of operators then actually coincides
with the von Neumann algebra they generate.) By construction, for any pair of
open intervals $I_1,\,I_2$ with $I_1\subseT I_2$ we have an inclusion
$\car (I_1) \subseT \car (I_2)$, which is
inherited from the natural embeddings of the $L^2$ spaces.
The global CAR algebra is defined as the norm closure
  \be \car:= \overline{\bigcup_{I\in\Js}\!\!\cari}  \Labl1C
of the union of local \alg s,
where $\Js$ denotes the set of those open intervals $I\subseT S^1$ whose
closure does not contain the point (say) $z\eq{-}1$.

This paper is organized as follows. In \srf F we summarize some features of
the DHR theory and study the interplay between the sector decompositions
that arise from embeddings of gauge groups. In \srf+ we present the sector
decompositions of the \ce\ `circle' \cfts\ with $2N$ sectors and of the
level one \sunh\ \wzwts\ and interpret them in terms of DHR gauge groups \zn.
The relation between these theories and the \cara s defined by \erf C is
demonstrated in \srf C. When the \zn\ gauge symmetry of these theories is
extended by a suitable further automorphism, one arrives at a description of the
\ce\ orbifold and the level two \sonh\ \wzwts\ with $N{+}7$ sectors; these
theories and their connection with the \cara\ is analyzed in \srf Q.
Finally, in \srf O we present the Fourier modes of the observables,
expressed in terms of the Fourier modes of the free fermions.

\Sect{DHR sectors and embeddings of gauge groups}F

Let us briefly recall some facts about the DHR theory \cite{dohr,dohr2,doro0}
of \suse s. We are dealing with chiral \cfts, so that the relevant space-time
is $S^1$, the unit circle. To apply the DHR theory to this situation,
one associates to each interval $I\subseT S^1$ a
{\em local field algebra\/} $\F(I)$; this is a von Neumann algebra which
acts on some Hilbert space $\H$ in such a way that
$I_1\subseT I_2$ implies $\F(I_1) \subseT \F(I_2)$. The {\em global\/}
(or {\em quasilocal\/}) field algebra, which is defined as the norm closure
  \be  \F = \overline{\bigcup_{I\in\Js}\!\!\F (I)}  \ee
of the union of all local field \alg s,
acts irreducibly on $\H$. Here as in \Erf1C $\Js$ denotes the set of open
intervals in $S^1$ whose closure does not contain ${-}1\iN S^1$.
The Hilbert space $\H$ carries a strongly continuous \rep\ $\VV$ of
the space-time symmetry group $\SUEE$ such that the generator $L_0$ of rotations
is positive and the eigenvalue zero belongs to a unique (up to a phase)
vacuum vector $\vac\iN\H$ (see e.g.\ \cite{frrs2,gafr}).
The field algebras transform covariantly
\wrt $\VV$. Furthermore, $\H$ also carries a strongly continuous \rep\ $U$ of a
compact group $G$, called the DHR {\em gauge group\/}, which commutes with
$\VV$ and leaves the vacuum vector invariant, and which transforms each local
field \alg\ into itself. {\em Local observable algebras} are the fixed
point \alg s of the field \alg s \wrtt gauge group $G$,
  \be \A (I) = \F (I)^G := \F (I) \cap U(G)' \,,  \ee
and the global observable \alg\ is
  \be \A = \overline{\bigcup_{I\in\Js}\!\!\A (I)} \,, \ee
so that $\A=\F^G$. Note that while all local \alg s are von Neumann
algebras, the global \alg s are only $C^*$-\alg s.
Fields are relatively local to the observables, and
this implies in particular locality of the observables.

Under these (and a few further standard) assumptions, the DHR theory tells
us that the Hilbert space \H\ decomposes as
  \be  \H=\oplusg\alpha \H_\alpha \otimes \complex^{d_\alpha}_\P  \,.  \Labl Ha
\wrt the action of \A.
Here $\H_\alpha$ are pairwise inequivalent irreducible \A-modules,
called the {\em superselection sectors}, $\hat G$ denotes the
group dual of $G$ (i.e.\ the \gdual\
of $G$, which constitutes a basis of the \rring\ of $G$), and $d_\alpha$ is the
dimension of the irreducible $G$-\rep\ $\pi_\alpha$ with character
$\alpha\iN\hat G$. The gauge group $G$ acts on the multiplicity space
$\complex^{d_\alpha}$ by the representation $\pi_\alpha$, i.e.
  \be  U(g)\eq\oplusgt\alpha \bfe_{\Hm_\alpha} \otim\,
  \pi_\alpha(g) \quad\ \mbox{for all} \;\ g\iN G \,. \ee

Next we investigate what happens when we are given two different DHR gauge
groups $G$ and $H$ which
act on one and the same \fielda\ \F. Then there are two decompositions
  \be  \oplusg\alpha \H_\alpha \otimes \complex^{d_\alpha}_\P = \H
  = \oplush a \H_a \otimes \complex^{d_a}_\P  \Labl GH
of the Hilbert space
\H\ \wrtt fixed point \alg s $\F\og$ and $\F\oh$, \resp. Now consider
the situation that $H\subseT G$ and that the action of $H$ is defined by the
restriction of $U$ from $G$ to $H$. It is not hard to see that the
decompositions \erf{GH} are then related as
  \be  \H_a =\oplusg\alpha \H_\alpha \otimes \complex^{n_\alpha^a}_\P  \,,
  \Labl aa
where the branching coefficients $n_\alpha^a$ are defined through
the restriction
  \be  \res GH{\pi_\alpha}= \oplush a n_\alpha^a\,\pi_a    \ee
of irreducible $G$-\rep s $\pi_\alpha$ to $H$-\rep s, or equivalently, through
  \be  \ind GH{\pi_a}\eq\oplusgt \alpha n_\alpha^a\,\pi_\alpha  \ee
by Frobenius reciprocity. In other words, the superselection sectors
$\H_a$, labelled by $a\iN \hat H$, are related to the sectors $\H_\alpha$,
$\alpha\iN \hat G$, according to the induction from $H$ to $G$.

We are particularly interested in the specific case where $H$ is embedded
diagonally into $G=H\Times H$. Then $\hat G=\hat H\Times\hat H$, so that
$\alpha\iN\hat G$ can be considered as a pair $(a_1{,}a_2)$ with $a_1,a_2
\iN\hat H$. It follows that
  \be  n_\alpha^a \equiv n_{(a_1,a_2)}^a = \N{a_1}{a_2}{\ \,a}  \,, \labl{resf}
where $\N abc$ are the {\em fusion coefficients\/} of $H$, defined by
the tensor product decomposition
  \be  \pi_a \otimes \pi_b \cong \oplush c \N ab{\!c} \, \pi_c \ee
of irreducible $H$-\rep s.
This observation applies in particular to the situation where the \fielda\ \F\
has the structure of a tensor product $\F\eq\FE\Otimes\FZ$ of \fielda s which
possess isomorphic DHR gauge groups, i.e.\ for which the associated \obsa s are
  \be  \AE=(\FE)\ohe\,,\quad \AZ=(\FZ)\ohz \qquad{\rm with}\;\
  \he\cong\hz\;{\cong:}\,H  \,.  \Labl12
In this case the \fielda\ \F\ acts in a canonical manner on the tensor
product $\H\eq\HE\otimes\HZ$ of Hilbert spaces \HE\ and \HZ, which
under the action of the \obsa s \AE\ and \AZ\ decompose into sectors as
  \be  \HE= \oplush a \hE_a \otimes \complex^{d_a}
  \qquad{\rm and}\qquad \HZ= \oplush a \hZ_a \otimes \complex^{d_a}\,, \ee
\resp. It follows from the result above that under the action of the diagonal
subgroup of $\he\Times\hz$, \H\ decomposes as
  \be  \H= \oplush a \H_a \otimes \complex^{d_a} \qquad{\rm with}\qquad
  \H_a= \oplush{b,c}\hE_b\otimes\hZ_c\otimes\complex^{\N bc{\!a}}_\P  \,.
  \Labl14

Below we will encounter the specific case of cyclic gauge group
$H\cong\zn\equiv\zet/2N\zet$. Then also $\hat H\cong\zn$; thus the labels $a\iN
\hat H$ can (and will) be considered as integers defined modulo $2N$, i.e.\
$\hat H\eq\{0,1\Ldots2N\mi1\}$, and the fusion coefficients read
$\N abc\eq\delta_{a+b,c}^{}$ for $a,b,c\iN\hat H$. Therefore the decomposition
\Erf14 reads $\H\eq\bigoplus_{a\in\zet_{2N}}\H_a$ with
  \be  \H_a= \oplusz b \hE_b\otimes\hZ_{a-b} \,;  \ee
in particular, the vacuum sector $\H_0$ splits as
  \be  \H_0= \oplusz a \hE_a\otimes\hZ_{2N-a} \,.  \ee

\Sect{$c\eq1$ and \wzwts\ with \zn\ fusion rules}+

For intervals $I\subseT S^1$ we denote by $\bosn(I)$ the local observable
\alg s of the \ce\ \cft\ with $2N$ sectors that corresponds to a (chiral) free
boson compactified on a circle of appropriate radius.
According to the results of \cite{bumt}, the \alg s $\bosn(I)$ are the von
Neumann \alg s that are generated by local bounded functions of a \ueh\ current
and of a conjugate pair of Virasoro-primary fields of
\cdim\ $\Delta\eq N$. Similarly, we denote by $\wzwu (I)$ the local
observable \alg s of the \wzwt\ based on the \sunh\ current \alg\ at level 1.
The corresponding global $C^*$-\alg s will be denoted by \bosn\ and
\wzwu, \resp, and the associated field algebras by $\Fa\,{\equiv}\,\FA$ and
$\Fs\,{\equiv}\,\FS$, \resp. Unfortunately, while explicit expressions for
localized fields are available in the \ce\ case \cite{bumt}, to the best of our
knowledge they are not known for the \wzwts. However,
both for the \ce\ and the \wzwts, `point-like localized' unbounded field
operators can be obtained by the vertex operator construction.
For the \ce\ theory the vertex operators are given by
  \be  \varphi_a(z)\Sim\normord{\eE^{\ii aX(z)/\sqrt{2N}}}  \Labl1v
for $z\iN S^1$, while for the WZW case they read
  \be  \psI_a(z)\Sim\normord{\eE^{\ii(\lI,Y(z))}}  \,.  \Labl2v
Here $a\iN\zn$, $X$ is a free boson and $Y\Equiv(Y^i_{})$
a collection of $2N\mi1$ free bosons, and
the colons stand for a suitable normal ordering prescription;
moreover, \li\ denote the fundamental weights of the \findim\ \lie\
\sun\ (for $a\Ne0$, together with $\lo\df0$), and $(\,\cdot\,,\cdot\,)$
is the inner product on the weight space of \sun.

Both of these two types of \qfts\ possess $2N$ sectors,
each of which has statistical dimension one, and in either case the
fusion ring is the group ring of the finite cyclic group \zn.
In this situation it is quite natural to expect that the
composition of sectors can be understood by promoting this \zn\ group
to a DHR gauge group; the following considerations demonstrate that this
is indeed the case. (Similar arguments will work for any other theory 
whose fusion ring coincides with the group ring of a finite abelian group.)

The sectors of the \ce\ and of the \wzwts\ can be obtained\,%
\futnote{To be precise, this actually yields only dense subspaces of the
sectors (the same remark applies to the sectors of the other theories treated
below). However, this will not play any \role\ in our discussion, and
accordingly we simplify notation by using the same symbols for the sectors
and for their dense subspaces.}
by applying the Fourier modes of the observables of these theories
to suitable \hw\ vectors; in both cases
it is in fact sufficient to employ only the generators of the relevant current
\alg s (which will be described in detail in \srf O). It follows that
the sectors of these theories are isomorphic, \resp, to direct sums\,%
\futnote{The individual summands are related to each other by the action of
the additional observables of \cwei\ $N$. For the present statements we do
not, however, need any information about the form of this action.}
of the \ihwm s $\Lu b$ of the \ueh\ current algebra with charge
$b\,{\rm mod}\,2N$, and to the \ihwm s $\Lll\equiv\LLl$ of the WZW theory with
certain specific highest \sun-weights $\Lambda$. More specifically, one finds
that the sectors $\Lb b\equiv\LB b$ of the \ce\ theory are the direct sums
  \be  \Lb b = \oplusZ n \Lu{b+2nN}  \,,  \ee
while the sectors $\Ll a$ of the level one \sun\ theory are modules
$\Lll$ whose \hw\ $\Lambda$ is either $\Lj0\eq0$ (for the vacuum sector) or a
fundamental weight $\Lambda_{(a)}$ ($a\iN\{\onetone\}$) of \sun.
Thus the sector decompositions of these theories read
  \be  \dhA = \oplusz a \Lb a \equiv \oplusz a \oplusZ n \Lu{a+2nN}
  = \oplusZ m \Lu m  \ee
and
  \be \dhS = \oplusz a \Ll a \,, \Labl ds
\resp.

{}From these decompositions we learn in particular that the spaces
$\dhA$ and $\dhS$ naturally carry \rep s of $\zn$. Since the group
$\zn$ does not possess a distinguished generator, there is a priori
some arbitrariness in the precise definition of these \rep s, though.
As it turns out, a convenient prescription for the
\rep s $\Ua\equiv\UA$ of $\zn$ on $\dhA$ and $\Us\equiv\US$ of \zn\
on $\dhS$ is provided by\,%
\futnote{We continue to use the additive notation for integers modulo $2N$.}
  \be  \Ua |_{\Lb a}:= \pi_a \qquad{\rm and}\qquad \Us |_{\Ll a}:= \pi_{-a}
  \,.  \labl{Urep}
Here for each $a\iN\zn$, $\pi_a$ denotes the \zn-\rep\ that acts as
  \be  \pi_a(b)= \eE^{\ii \pi ab /N} \quad\ {\rm for\ all}\;\ b\iN\zn \,.  \ee

\Sect{\cara s from \cfts\ with \zn\ fusion rules}C

The discussion of the previous section shows in particular that we are dealing
with a situation of the type described in equation \Erf12 above, with\,%
\futnote{Here and in the following, equalities and isomorphisms
between \alg s are meant to apply both to global and to local algebras.}
  \be  \bosn \equiv \AA = (\FA)\oz \,,\qquad
  \wzwu \equiv \AS= (\FS)\oz \,.  \ee
It is then natural to ask what the fixed point \alg\ of $\F\equiv\FA\Otimes\FS$
\wrtt diagonal subgroup $\znd\subset\zn\Times\zn$ looks like.
As usual, $\F$ acts irreducibly on a Hilbert space $\dh$, and under
the action of $\F\ozd$ we have a decomposition $\dh = \Oplusz a\dh_a$.
As it turns out, $\F\ozd$ is nothing but the \cara\ \erf C
for $2N$ complex fermions, i.e.
  \be  (\FA\Otimes\FS)\ozd \cong \car  \,,  \labl c
where it is understood that the action of $(\Fa\Otimes\Fs)\ozd$ is restricted
to the vacuum sector $\dh_0$ \wrtt diagonal $\zn$ subgroup.
Note that \erf c implies in particular that
  \be  \bearll  \car\oz \!\!& \cong (\FA\Otimes\FS)\oztz \\[.6em]&
  = (\FA)\oz \otimes (\FS)\oz = \bosn \otimes \wzwu  \,.  \eear \Labl cz

We will study this relationship in \srf O in terms of the Fourier modes of
the observables. Here we verify \erf c in terms of the superselection sectors,
i.e.\ show that the vacuum sector $\dh_0$ on which $(\Fa\Otimes\Fs)\ozd$
is acting coincides with \fock.
Recall from the introduction that we can construct the sectors $\dcar_a$ of
the fermion theory by applying the Fourier modes of the fermions to the vacuum.
Now the results on the Hilbert spaces of the \ce\ and \wzwts\ that we listed in
the previous section imply in particular that
(a dense subspace of) the Hilbert space of the tensor product theory is
  \be  \dh = \dhA \Otimes\dhS \cong
  \oplusz{a,b}\!\! \LLb \Lb a \otimes \Ll b  \LRb
  = \oplusz{a,b}\!\! \LLb\lLb \oplusZ n \Lu {a+2nN} \lRb \otimes \Ll b \LRb
  \,. \Labl da
Furthermore, it is known \cite{hase} that the space
$\dcar\eq\Oplusz a\dcar_a$ decomposes as
  \be  \dcar \cong \oplusz b  \LLb \lLb \oplusZ n \Lu{b+2nN}
  \lRb  \otimes \Ll b  \LRb   \Labl dc
into a direct sum of tensor products of the current \alg\ modules that
appear in the former decompositions.

We would like to compare the results \Erf da and \Erf dc from the perspective
of the DHR situation that we studied in \srf F. To this end we
employ the \zn-\rep s $\Ua$ and $\Us$ \erf{Urep}.
On $\dh$, these \rep s induce an action of $\zn\Times\zn$ according to
  \be  \zn\Times\zn \ni (a,b)\,\mapsto\, \Ua(a)\,\otim\,\Us(b) \,.  \ee
We can then restrict this \rep\ to the diagonal subgroup of $\zn\Times\zn$
and decompose $\dh$ into its sectors $\dh_a$ \wrt this diagonal subgroup:
  \be  \dh = \oplusz a \dh_a  \quad \ {\rm with}\quad
  \dh_a := \oplusz b\!\! \LLb\lLb \oplusZ n \Lu{a+b+2nN}\lRb \otimes \Ll b \LRb
  \,. \Labl dA
Comparison of \Erf da with \Erf dc now tells us that
  \be  \dcar \cong \oplusz a \Lb a \otimes \Ll a = \dh_0\,. \Labl dd
This finally confirms the validity of the isomorphism \erf{c}.

Let us also mention that when we consider the action of the
full group $\zn\Times\zn$ instead of its diagonal subgroup, then we must
have an additional \zn\ action in the fermionic Fock space
$\hcar$ of the $2N$ complex fermions. To fit with the previous results,
for each $a\iN\zn$ such an additional gauge transformation must act
on the fermion modes $\b ir$ and $\c ir$ as the Bogoliubov transformation
  \be  \bogzn a:\quad \b ir \mapsto \eE^{\ii\pi a/N} \b ir \,,
  \qquad  \c ir \mapsto \eE^{-\ii\pi a/N} \c ir \,. \Labl zf

It is illustrative to formulate the results above also in terms of
the (Virasoro-specialized) characters
  \be  \chii_V(q):={\rm tr}^{}_V\;q_{}^{L_0} \ee
of the various vector spaces $V$ that appeared in the decompositions.
We first note that \Erf zf can be regarded as the restriction of an action
of the gauge group $\Ue$, for which $a\iN\zn$ is just to be
replaced by an arbitrary real parameter. The characters of the sectors
of the \alg\ $\car\ozd$ can therefore be calculated by first considering
the decomposition of $\hcar$ into sectors $\H_m\oUe$, $m\iN\zet$, \wrt
a gauge group $G\eq\Ue$ and then the restriction of $G$ to its subgroup
$H\eq\zn$. In this situation we can again apply the relation \Erf aa, which
tells us that
  \be  \H_a =\oplusZ n \H_{a+2nN}\oUe  \ee
for each $a\iN\zn$.

Now the characters for the spaces $\H_m\oUe$ read (see e.g.\ \ci)
  \be  \chii_m\oUe(q) = \sN m \, (\vi(q))^{-2N} \,, \ee
while the characters of the sectors $\dha_a$ of \Fa\ and
$\dhs_a$ of \Fs\ are given by
  \be  \chiA_a(q) = (\vi(q))^{-1}\, \psif Na \ee
and by
  \be  \chiS_a(q)= q^{-a^2/2N}\,(\vi(q))^{1-2N}\,\sN a \,, \ee
\resp. Here
  \be  \vi(q) := \prod_{n=1}^\infty (1-q^n) \Labl vi
is Euler's product function, while
  \be  \sN a := \!\! \summN a q_{}^{(m_1^2+m_2^2+...+m_{2N}^2)/2}  \Labl sN
and
  \be  \psif Na:= \sum_{m\in\zet} q^{(a+2mN)^2/4N}  \labl{psif}
for $a\iN\zn$.
It is then easy to verify (compare e.g.\ formula (9.17) of \ci) that indeed
  \be  \chii_a(q) = \sumz n \chii_{a+2nN}\oUe(q) =
  \chiA_a(q) \cdot \chiS_a(q)  \ee
for all $a\iN\zn$, in agreement with our result that
the sectors of the $\zn$ fermion algebra are precisely the
tensor products of the irreducible modules for $\bosn$ and $\wzwu$.

As another consistency check of the relation \erf c
we verify that the fields in the \alg\ $(\FA\Otimes\FS)\ozd$ possess the
right braiding properties. Since
the form of the braid relations does not depend on the precise choice of
localization of the fields, we can consider the `point-like localized'
unbounded field operators \Erf1v and \Erf2v. These vertex operators
carry an abelian (`anyonic') \rep\ of the braid group, and the phases appearing
in this \rep\ can be determined from the \cwei s of the vertex operators.
More specifically, for the \ce\ theory we have $\cd{\varphi_a}\eq a^2/4N$, and
accordingly \cite{bumt} $\varphi_a(z)\,\varphi_b(w)= \eE^{\ii\pi\epsilon\,ab/2N}
\varphi_b(w)\,\varphi_a(z)$ (with the sign $\epsilon\iN\{\pm1\}$ in the
exponent depending on whether $w$ is to the `left' or to the `right' of $z$ on
the punctured circle), while for the \wzwt\ we have $\cd{\psI_a}\eq a(2N-a)/4N$
so that $\psI_a(z)\,\psI_b(w)=\pm\eE^{-\ii\pi\epsilon\,ab/2N} \psI_b(w)\,\psI_a
(z)$. Those fields in the tensor product theory which are invariant under the
diagonal \zn\ gauge group therefore all have \cwei\ $a/2$ for some $a\iN\zn$ and
satisfy fermionic or bosonic braiding relations, as is required for \erf c to
hold.

\Sect{Theories with gauge group $\dn$}Q

It is known that the sectors of the \ce\ circle theory with \obsa\ \bosn\
combine, respectively decompose, into the untwisted sectors of the \ce\ 
orbifold theory that has $N\pl7$ sectors \cite{dvvv}. Moreover, inspection
of the results of \cite{dvvv} also shows that
these untwisted orbifold sectors generate a fusion ring which is isomorphic
to the representation ring of the {\em generalized quaternion group\/}
\dn\ (see appendix A for some basic information about these non-abelian
finite groups). Similarly, the \wzwu\ sectors give rise to
the untwisted sectors of the level two \sonh\ \wzwt, which can be seen to 
generate a \dn\ fusion ring as well. These observations lead us to expect that,
as far as the untwisted sectors are concerned, the superselection structure of
the \ce\ orbifold and level two \sonh\ theories can be understood in terms of a
DHR gauge group \dn.
In this section we show that indeed one obtains the observable
algebras of these theories when one extends the \zn\ gauge groups that appeared
in the previous setting to \dn. Precisely speaking, we claim that we have
an action of \dn\ such that
  \be  (\FA)\od = \orbn \qquad{\rm and}\qquad (\FS)\od = \wzwo \,,  \labl{claim}
where \orbn\ and \wzwo\ stand for the \obsa s of the \ce\ orbifold theory
with $N\pl7$ sectors and of the level two \sonh\ \wzwt, \resp.

To prove this claim and to study its consequences, it is convenient to
express the action of the gauge groups in terms of automorphisms of the
relevant field algebras. For each $a\iN\zn$ the representations $\UE\equiv\UA$
and $\UZ\equiv\US$ that were introduced
in \erf{Urep} define automorphisms $\Ad{\Uj(a)}$
of the field algebras $\Fj$ for $j\eq1,2$, respectively. We denote by $x$ the
generator of the abstract group $\zn$ ($x^{2N}\eq1$), and define
  \be  \ax\indj\df\Ad{\Uj(x)}\in\Autj  \ee
for $j\iN\{1,2\}$. We wish to consider
the situation that the \zn\ gauge groups of both the \ce\ and the \wzwt\ get
extended to the generalized quaternion groups \dn, by including another
generator $y$ in such a way that the relations \Erf3a of \dn\ are satisfied.
This means that for $j\eq1,2$ we have besides $\ax\indj$
automorphisms $\ae\indj\eq\Ad{\Uj(y)}\iN\Autj$ which obey
  \be  \ax\indj^{2N}=\id \,,\qquad
  \ax\indj \circ \ae\indj \circ \ax\indj = \ae\indj \,, \qquad
  \ae\indj^2 = \ax\indj^N \,.  \labl{autq}

As a first consequence of our claim we observe that
it gives rise to the identifications
 \be   (\bosn)\ozz = \orbn \qquad{\rm and} \qquad (\wzwu)\ozz = \wzwo   \ee
for a suitable $\zz$ group of \auto s.
This can be seen as follows. By the identities \erf{autq} we have
$\ax\indj\Circ\ae\indj(F\indj)\eq\ae\indj\circ\ax\indj^{-1}(F\indj)$
for all $F\indj\iN\Fj$.
As a consequence, $A\indj\iN(\Fj)\oz$ implies that also
$\ae\indj(A\indj)\iN(\Fj)\oz$, and hence there exist restrictions $\bogzz\indj$
of $\ae\indj$ to the $\zn$-invariant subalgebras $\bosn\eq(\FE)\oz$
and $\wzwu\eq(\FZ)\oz$, \resp. Now by definition the
automorphism $\ax\indj$ of $\Fj$ restricts to the identity on $(\Fj)\oz$, and
therefore the relations \erf{autq} imply that $\bogzz\indj^2\eq\id\,$ for
$j\eq1,2$. Thus $\bogzz\indj\iN\Aut{(\Fj)\oz}$
are in fact $\zz$-automorphisms. Put differently,
the restrictions $\bogzz\indj$ exist because $\zn\subseT\dn$ is a normal
subgroup, and they are $\zz$-automorphisms because $\dn/\zn \cong \zz$.

Now let us consider the automorphism
  \be  \ax:=\ax\inde\otiM\ax\indz
  \,\in\Aut \F \equiv \Aut {\FE\Otimes\FZ} \,, \ee
which represents the generator $(x,x)$ of the diagonal
subgroup $\znd\subseT\zn\Times\zn$. The action of the full group
$\zn\Times\zn$ can be obtained by including another \zn-automorphism, say
  \be  \axt:=\id\otiM\ax\indz \,,  \ee
which realizes the element $(1,x)$ of $\zn\Times\zn$. Now of course $\ax$ and
$\axt$ commute (or in more mathematical terms, the diagonal subgroup
$\znd\subseT\zn\Times\zn$ is normal), so that there exists a restriction $\az$
of $\axt\iN\Aut\F$ to $\F\ozd$. We claim that when evaluated on the \cara\,%
\futnote{Recall that in this isomorphism it is understood that the action of
$\F\ozd$ is restricted to the vacuum sector \wrt \znd.}
$\car\cong\F\ozd$, this restriction coincides with the Bogoliubov
automorphism $\bogzn1$ that was defined in formula \erf{zf}:
  \be  \az=\bogzn1:\quad \b ir \mapsto \eE^{\ii\pi/N}\b ir \,,
  \quad \c ir \mapsto \eE^{-\ii\pi/N}\c ir \,.  \Labl az
Next we also define $\ae:=\ae\inde\otiM\ae\indz\iN\Aut{\FE\Otimes\FZ}$.
Clearly, $\ax$ and $\ae$ generate the diagonal subgroup
$\dnd\subset\dn\Times\dn$. Now denote by $\bogzz$ the restriction of
$\ae\iN\Aut\F$ to $\F\ozd$, which again exists because $\znd\subseT\dnd$
is normal. Since $\ax$ acts trivially on $\F\ozd$, we learn that
$\bogzz^2\eq\id$, i.e.\ $\bogzz$ generates a \zz\ group. We claim
that this \zz-action on $\car\cong\F\ozd$ is realized as the exchange
  \be  \bogzz:\quad \b ir \,\leftrightarrow\, \c ir   \Labl zz
of the fermion modes.

We will verify our claims in the following analysis of characters. Moreover,
we will see in \srf O that the $\bogzz$-invariant linear combinations
of \wzwu-generators are precisely the generators of \wzwo, and that the
$\bogzz$-invariant combinations of \bosn-generators are precisely the
generators of \orbn. We start by noting
that $[\dn\,{:}\,\zn]=2$. Correspondingly it is not difficult to
construct the induction of the \irrep s $\pi_a$ of \zn\ to its extension \dn.
Namely, for $0{<}a{<}N$ the irreducible \zn-\rep s $\pi_a$ and $\pi_{2N-a}$
combine to a \twodim\ \irrep\ of \dn,
while $\pi_0$ and $\pi_N$ split into the direct sum of two \onedim\ \rep s.
Denoting the group dual (i.e., the \gdual) of the non-abelian group \dn\ by
  \be  \dnh = \{\dno,\dnv,\dns,\dnc\} \cup \{a\,|\,a\eq\onetond\} \,, \ee
this is written as
  \be \bearl  \ind\Dn\Zn{\pi_a}= \ind\Dn\Zn{\pi_{2N-a}}
  = \pid_a \qquad{\rm for}\,\ 0\,{<}\,a\,{<}\,N \,, \\[.95em]
  \ind\Dn\Zn{\pi_0^{}} =\pido\oplus\pidv\,,
  \qquad \ind\Dn\Zn{\pi_N^{}} = \pids\oplus\pidc  \,.  \eear \labl{indZQ}

Next we study what happens when we extend the diagonal subgroup
$\znd\subseT\zn\Times\zn$ to the diagonal subgroup $\dnd\subseT\dn\Times\dn$.
Applying the general result \erf{aa} to the situation described by \erf{indZQ},
we learn that under the action of $\F\odd\subseT\F\ozd$
the fermionic Fock space $\hcar\equiv\dh_0$ splits as
$\dh_0=\dh_\dno \oplus \dh_\dnv$. Moreover,
$\dh_\dno=\hcarp$ and $\dh_\dnv=\hcarm$ are the even and odd
subspaces with respect to the $\zz$-automorphism $\bogzz$ of \erf{zz},
respectively. When we further extend the gauge group to the full $\dn\Times\dn$
group, these spaces further decompose into sectors of
$\F\odtd\eq(\Fa)\od\Otimes(\Fs)\od$. Now the latter are of course
tensor products of the sectors of $(\Fa)\od$ and $(\Fs)\od$, which we
denote by $\Ha_\alpha\Equiv\hA_\alpha$ and $\Hs_\beta\Equiv\hS_\beta$,
respectively. Implementing once again \erf{aa} as well as the formula
\erf{resf}, we then conclude that this decomposition reads
  \be  \hcarp = \Oplusd{\alpha,\beta} \Ha_\alpha \otimes \Hs_\beta
  \otimes \complex_{}^{\N\alpha\beta\dno} \,, \qquad
  \hcarm = \Oplusd{\alpha,\beta} \Ha_\alpha \otimes \Hs_\beta
  \otimes \complex_{}^{\N\alpha\beta\dnv} \,, \ee
where $\N\alpha\beta\gamma$ are the fusion coefficients of \dn, i.e.\ more
explicitly,
  \be \bearll
  \hcarp \!\!& = \Ha_\dno\OTimes\Hs_\dno \oplus \Ha_\dnv\OTimes\Hs_\dnv
  \oplus\dstyle\bigoplus_{\alpha=1}^{N-1}\! \Ha_\alpha\OTimes\Hs_\alpha
  \\{}\\[-.98em] & \hsp{6.13}
    \oplus \left\{ \bearll \Ha_\dns\OTimes\Hs_\dns \oplus
    \Ha_\dnc\OTimes\Hs_\dnc & \quad \mbox{for $N$ even,} \\[.23em]
    \Ha_\dns\OTimes\Hs_\dnc \oplus
    \Ha_\dnc\OTimes\Hs_\dns & \quad \mbox{for $N$ odd,} \eear \right.
  \\{}\\[-.5em]
  \hcarm \!\!& = \Ha_\dno\OTimes\Hs_\dnv \oplus \Ha_\dnv\OTimes\Hs_\dno
  \oplus\dstyle\bigoplus_{\alpha=1}^{N-1}\! \Ha_\alpha\OTimes\Hs_\alpha
  \\{}\\[-.91em] & \hsp{6.13}
    \oplus \left\{ \bearll \Ha_\dns\OTimes\Hs_\dnc \oplus
    \Ha_\dnc\OTimes\Hs_\dns & \quad \mbox{for $N$ even,} \\[.23em]
    \Ha_\dns\OTimes\Hs_\dns \oplus
    \Ha_\dnc\OTimes\Hs_\dnc & \quad \mbox{for $N$ odd.} \eear \right.
  \eear \Labl+-
On the other hand, it is not difficult to check -- some details are provided in
appendix B -- that the characters of $\hcarp$ and $\hcarm$ (i.e.\ the characters
of the \irmod s of $\car^\Zz$) can be decomposed as
  \be  \bearl
  \chip = \chic_\dno\, \chiz_\dno + \chic_\dnv\, \chiz_\dnv
    + \chic_\dns\, \chiz_\dns + \chic_\dnc\, \chiz_\dnc
    +\dstyle\sum_{\alpha=1}^{N-1} \chic_\alpha\, \chiz_\alpha \,,
  \\{}\\[-.76em]
  \chim = \chic_\dno\, \chiz_\dnv + \chic_\dnv\, \chiz_\dno
    + \chic_\dns\, \chiz_\dns + \chic_\dnc\, \chiz_\dnc
    +\dstyle\sum_{\alpha=1}^{N-1} \chic_\alpha\, \chiz_\alpha \eear \labl{evod}
into products of irreducible characters $\chic$ of the \ce\ orbifold theory
\orbn\ and of irreducible characters $\chiz$ of the level two \sonh\ \wzwt.
Now comparing the formulae \erf{+-} and \erf{evod} (and recalling that
$\chic_\dns\eq\chic_\dnc$ as well as $\chiz_\dns\eq\chiz_\dnc$) leads us
to the conclusion that the
sectors of the gauge invariant algebras $(\Fa)\od$ and $(\Fs)\od$ indeed
coincide with those of $\orbn$ and $\wzwo$, respectively. This
finally reproduces the statement of our claim.

We further support
our claim by the following consideration. Recall that $\bogzz\iN\Aut{\F\oz}$
satisfies $\bogzz^2=\id$. This implies that $\az$ and $\bogzz$ fulfill
  \be  \az^{2N}=\id \,,\qquad \az \circ \bogzz \circ \az = \bogzz \,, \qquad
  \bogzz^2 = \id \,.  \ee
Thus they provide a representation of the dihedral group $\dihed =
\zn \semitimesr \zz$ by automorphisms of $\F\ozd$. We can therefore consider
the \alg\ $(\F\oz)\odihed$, which is the invariant part of $\F\ozd$ with
respect to the automorphisms $\axt$ and $\ae$, or what is the same,
the invariant part of \F\ with
respect to $\ax$, $\axt$ and $\ae$. Now $\ax$, $\axt$ and $\ae$ provide a
representation of a subgroup $K_N\subseT\dn\Times\dn$ that is obtained by
adjoining the diagonal generator
$(y,y)$ to $\zn\Times\zn$. Note that $[\dn\Times\dn\,{:}\,K_N]=2$ and that
the diagonal subgroup $\znd\subseT K_N$ is normal. Now $K_N / \znd\cong\dihed$,
and this is the reason why the restrictions $\az$ and $\bogzz$ of
$\axt$ and $\ae$ provide a representation of $\dihed$ in $\Aut {\F\oz}$.
We conclude that
  \be  (\FA\Otimes\FS)\odtd \subset \F\ok \equiv (\F\oz)\odihed  \,,  \ee
or in other words,
  \be  \orbn\Otimes\wzwo \subset \car\odihed \,.  \labl{sub}
As it is a rather tedious calculation, we refrain from applying the
whole machinery of \srf F to the (gauge) subgroup
$K_N\eq H\subset G\eq \dn{\times}\dn$. Rather, we restrict the discussion to
confirming the validity of the \zz\,-type inclusion \erf{sub}, which is achieved
by the following argument. Under the action of $\car\odihed \cong \F\ok$
the vacuum sector $\dh_0\eq\hcar$ of $\F\ozd$ splits into sectors labelled by
$\dihedh$, and as the inclusion \erf{sub} is of \zz-type, in this process
each sector can split into at most two $\F\odtd$-sectors.
Now the group \Dihed\ is precisely represented in $\Aut \car$ by the
automorphisms \erf{az} and \erf{zz}, while for the characters of $\car\odihed$
we obtain
  \be  \bearl
  \chido= \chic_\dno \chiz_\dno + \chic_\dnv \chiz_\dnv \,, \qquad
  \chidv= \chic_\dno \chiz_\dnv + \chic_\dnv \chiz_\dno \,,\\[.7em]
  \chids= \chic_\dns \chiz_\dns + \chic_\dnc \chiz_\dnc \,, \qquad
  \chidc= \chic_\dns \chiz_\dns + \chic_\dnc \chiz_\dnc \,,\\[.7em]
  \chid\alpha=\chic_\alpha \chiz_\alpha \qquad{\rm for}\ \alpha\iN\{\onetond\}
  \,.  \eear \labl{mess}
(for more details, see appendix B).
Again we conclude by comparison that \erf{mess} precisely corresponds
to the decomposition of $\car\odihed$-sectors into tensor products of the
sectors of \orbn\ and \wzwo.\,%
\futnote{Another indirect confirmation of our claim follows from the
following observation. In the space
$\hcar$ there must be simultaneous highest weight vectors of $\orbn$ and $\wzwo$
which correspond to the blocks that appear in the decomposition \erf{mess}. In
\cite{bofu} we determined the larger set of simultaneous highest weight vectors
of $\wzwo$ and the orbifold Virasoro algebra. Now inspecting the
orbifold conformal weights of those vectors one learns that the simultaneous
highest weight vectors of $\orbn$ and $\wzwo$ are given by the zero grade
($n\eq0$) vectors among those
in Eqs.\ (8.8), (8.11), (8.12), (8.13) and also (10.12) and (10.13) of
\cite{bofu}. One can check that the $\dihed$ transformation
properties of these vectors are indeed in agreement with the
sector decomposition \erf{mess}.}

The above arguments in favor of our claims are certainly not rigorous,
because we employ the Virasoro-specialized characters which do not
encode the complete structure of the respective vector spaces. Nevertheless
our claims are bound to be correct. For instance,
it would otherwise be a complete mystery why relations among characters of the
type derived above should be valid. Note in particular that our
formulae hold simultaneously for all values of the integer $N$.
Moreover, the quite different arguments that we will
present in the following section provide further support to our claims.

\Sect{Fourier modes of the observables}O

In this section we express the Fourier modes of the observables for the
various models of our interest through the Fourier modes of the free fermions.
This will in particular allow us to confirm
various statements made earlier from a different point of view.

We first need to define a normal ordering of bilinears of fermion modes.
We adopt the convention that the symbol $\bc ir$ ($r\iN\zet{+}1/2$,
$i\iN\{\oneton\}$), stands for either of the Fourier modes $\b ir$ or $\c ir$
of the fermions. Our normal ordering prescription then reads
  \be  \normord{\bc ir\bc js} := \left\{ \bearll
  \bc ir\bc js & {\rm for}\ s>0\,,\\[.3em] -\bc js\bc ir & {\rm for}\ s<0 \,.
  \eear\right. \Labl no

One now checks by direct computation that the combinations
  \be  \J ijm:=\sumzh r \normord{\b ir\c j{m-r}}  \labl J
with $i,j\iN\{\oneton\}$ and $m\iN\zet$ satisfy the commutation relations
  \be  [\J ijm,\J kln] = \delta_{jk}^{}\J il{m+n} - \delta_{il}^{}\J kj{m+n}
  + m\,\delta_{jk}^{}\delta_{il}^{}\,\delta_{m+n,0}^{}\,\bfe \Labl53
and hence span a level one \glnh\ current \alg.
More precisely, this \lie\ is the direct sum of a level one \sunh\ affine \kma\
and a \ueh\ current \alg. The generators of the level one \sunh\ \alg\
consist of the linear combinations
  \be  H^i_m:=\J iim-\J{i+1}{\,i+1}m \qquad{\rm for}\ i\eq\onetone  \labl H
and of
  \be  E^{ij}_m:= \J ijm  \qquad{\rm for}\ i\ne j   \,.  \labl E
The zero modes ($m\eq0$) generate a subalgebra isomorphic to the simple \lie\
\sun. In particular, the modes $H^i_0$ span the Cartan subalgebra of \sun,
and for $i\,{<}\,j$ the $E^{ij}_0$ constitute the raising operators of \sun,
corresponding to the
positive \sun-roots $\alpha_{ij}\equiv\alpha_{(i)} +\alpha_{(i+1)} +\cdots+
\alpha_{(j-2)}+\alpha_{(j-1)}$ (where $\alpha_{(k)}$ denote the simple roots
of \sun), while for $j\,{<}\,i$ they are lowering operators, corresponding to
the negative \sun-roots $-\alpha_{ji}$. The \ueh\ current \alg\ is spanned by
  \be  \je m:=\Frac1{\sqrt{2N}}\sumi \J iim  \,, \Labl je
with relations
  \be  [\je m,\je n]=\delta_{m+n,0}^{}\,\bfe \,.  \ee

The *-operation acts on the currents \erf J as $(\J ijm)^* =\J ji{-m}$,
and their commutation relations with the fermion modes read
  \be  [\J ijm,\b kr] = \Del jk\, \b i{m+r} \,, \qquad
  [\J ijm,\c kr] = -\Del ik \, \c j{m+r} \,,  \Labl Ja
so that in particular
  \be  [\je m,\b ir] = \Frac1{\sqrt{2N}}\, \b i{m+r} \,, \qquad
  [\je m,\c ir] = -\Frac1{\sqrt{2N}}\, \c i{m+r} \,. \ee
In addition to the current \alg, the free fermions also bring along a
Virasoro algebra, with generators
  \be  L_m^\caR =-\onehalf\sumzh{r}\ \sumI\,
  (r-\Frac m2)\,\normord{\b ir\c i{m-r}} \,.  \Labl lc

Let us now interpret these observations from the perspective of sections
\ref{s.+} and \ref{s.C}. The (unbounded)
observables of the \sun\ level one \wzwt\ are well known \cite{knza}.
They consist precisely of the \sunh\ currents with modes \erf H and \erf E,
together with fields that are obtainable from the currents by taking
derivatives and forming normal-ordered products. Among the latter there is
in particular the associated Sugawara \emt, the Fourier modes
of which are given by the affine Sugawara formula
  \be  L_m\Inds = \sumz n \lLb
  \sumN{\scs i,j=1 \atop \scs i<j}\normord{\J ijn\J ji{m-n}} + \half\sumn{i,j}
  G_{ij}\,\normord{H^i_n H^j_{m-n}} \lRb  \Labl51
($G_{ij}$ denotes the inverse of the Cartan matrix of \sun, and the
normal ordering prescription is similar to the one in \Erf no), and hence
they lie in a suitable completion of the universal enveloping \alg\ of the
current modes.

It follows that in order to be in agreement with the isomorphism \Erf cz,
the observable algebras $\bosn(I)$ of the \ce\ theory are given by the
commutants of $\wzwu$ in $\car(I)\oz$. This certainly includes the
bounded local functions of the \ueh\ current $J$\,%
\futnote{Thus in particular the currents associated to $\Fa\Otimes\Fs$ are
represented
in the form of a tensor product, i.e.\ the \ueh\ current acts as $J\otim\bfe$
while the \sunh\ currents act as $\bfe\otim H^i_m$ and $\bfe\otim E^{ij}_m$.}
as well as those of the associated \emt\ whose modes are
  \be  L_m\Inda = \half\sumz n \normord{\je n\je{m-n}} \,.  \labl{lbos}
(By carefully treating multiple normal orderings of the fermion modes,
one can check that $L_m\Inda+L_m\Inds=L_m^\caR$.)
We are now looking for further unbounded observables associated to the
$\bosn$ theory. In particular, we would like to find expressions that
commute with the $\sunh$ modes and can be interpreted as the Fourier modes of
fields $\phi$ that have integral \cwei\ $\cd\phi$ (\wrt $L_0^\caR$) and that
can play the r\^ole of primary conformal fields in the sense of \cite{bepz}.
Then the associated local bounded functions should be elements of $\bosn$.
Taking into account the isomorphism \erf{cz}, we have to expect that these modes
are $\zn$-invariant infinite series of normal-ordered products of fermion modes;
moreover, we in fact need only to
consider normal-ordered multilinears which are summed over like
  \be  \phi^{i_1i_2...i_p}_m = \sumzH{1.1}{r_1,r_2,...,r_{p-1}}
  \normord{\bc{i_1}{r_1}\bc{i_2}{r_2}\cdots
  \bc{i_{p-1}}{r_{p-1}} \bc{i_p}{m-r_1-r_2-...-r_{p-1}}}   \Labl pm
with $p\eq2\cd\phi$ even. In addition,
we can focus our attention to a small subset of these unbounded observables,
namely to those from which all others can be obtained by the operations of
taking derivatives and of forming normal-ordered products; for brevity,
we will refer to this subset as the {\em basic\/} (unbounded) observables.

Now by a suitable relabelling of the summation indices on the right-hand side
of \Erf pm we deduce from the anticommutativity of the $b$ \resp\ the $c$ modes
among themselves that we can assume that equality $i_k\eq i_l$ for $k\Ne l$
appears only if one is dealing with two different types of modes, i.e.\ only if
$\bc{i_k}{r_k}\eq\b{i_k}{r_k}$ and $\bc{i_l}{r_l}\eq\c{i_l}{r_l}$ (or the
other way round).
Employing the basic commutation relations \Erf Ja to compute the commutator of
$\phi^{i_1i_2...i_p}_m$ with $\J jkn$ one then finds the following.
First note that expressions which are neutral \wrtt gauge group \zn\ \Erf zf
must involve products for which the numbers of $b$ and $c$ factors differ by
a multiple of $2N$.  Now when \Erf pm contains an equal number of $b$ and
$c$ modes, then it commutes with the \sunh\ currents
precisely if it is a normal-ordered product of the Fourier modes of the
\ueh\ current $\Je$ (i.e.\ the $\je m$ \Erf je) and/or its derivatives.
A similar analysis shows that for any other \zn-neutral combination $X$
to commute with the \sunh\ currents it is necessary and sufficient that $X$
is obtainable by taking derivatives and/or forming normal-ordered products
of the \ueh\ current modes and of the modes\,%
\futnote{The summations ensure locality. Also note that owing to
$[\b ir,\b js]=0=[\c ir,\c js]$ no normal ordering is required here.}
  \be  \bearl
  \B m := \sumZH{1.5}{r_1,r_2,...,r_{2N-1}} {\b1{r_1}\b2{r_2}\cdots
  \b{2N-1}{r_{2N-1}} \b{2N}{m-r_1-r_2-...-r_{2N-1}}} \,, \\{}\\[-.9em]
  \C m := \sumZH{1.5}{r_1,r_2,...,r_{2N-1}} {\c1{r_1}\c2{r_2}\cdots
  \c{2N-1}{r_{2N-1}} \c{2N}{m-r_1-r_2-...-r_{2N-1}}} \,. \eear \Labl BC
One can check that the point-like localized fields which have \Erf BC as their
Fourier modes, namely $B(z)\df\sum_{m\in\zet}z^{m-N}\B m\eq{\b1{}(z)
\b2{}(z)\cdots\b{2N}{}(z)}$ and $C(z)\df\sum_{m\in\zet}z^{m-N}\C m
\eq\c1{}(z)$\linebreak[0]$\c2{}(z)\cdots\c{2N}{}(z)$, are primary conformal
fields and have \cwei\ $\cd B\eq\cd C\eq N$, both \wrtt fermion \emt\ \Erf lc
and \wrtt \ce\ \emt\ \erf{lbos} (which is compatible because they commute with
the \sunh\ current \alg\ and hence with its \emt).

We conclude that the basic observables of the \ce\ theory consist of the
\ueh\ current and the fields $B$ and $C$, which are related by charge
conjugation. This is in complete agreement with
the description of the observables that was given in \cite{bumt}.
The commutation relations of the modes \Erf BC are of the form
  \be  \bearl  [\B m,\B n]= 0 = [\C m,\C n] \,,  \\{}\\[-.8em]
  [\B m,\C n]= \coef1\,\delta_{m+n,0}\,\bfe - \coef2\,\jet{m+n}
  + \half\coef3\,\lLb\normord{\Jet^2}-\partial\Jet\lRb_{m+n} + \ldots \,,
  \eear \Labl bc
where
  \be  \coef j \equiv \coef{m,N;j} := \Frac1{(2N-j)!}\,\sum_{\ell=1}^{2N-j}
  (m-N+j+\ell-1)  \ee
for $j\iN\natnum$ and
  \be  \jet m:= \sqrt{2N}\,\je m \,,  \ee
and where the ellipsis stands for further terms involving $j$-fold
normal-ordered products of the currents (combined with terms involving
derivatives, in a similar way as for $\normord{\Jet^2}-\partial\Jet$\,)
and coefficients whose $m$- and $N$-dependence takes the form of $\coef j$,
for $j\eq4,5,...\,,2N\mi1$.

Note that for $N\eq1$, the relations \Erf bc amount to the statement
that besides the \suzh\ \aff\ with generators \erf H and \erf E, i.e.
  \be  \bearl
  J^+_m := E^{12}_m = \sumzh r \normord{\b 1r \c 2{m-r}} \,, \qquad\
  J^-_m := E^{21}_m = \sumzh r \normord{\b 2r \c 1{m-r}} \,, \\{}\\[-.8em]
  J^0_m := H^1_m = \sumzh r (\normord{\b 1r \c 1{m-r}}
  - \normord{\b 2r \c 2{m-r}}) \,, \eear  \Labl18
another (relatively commuting) level one \suzh\ \alg\ is present, namely
the one generated by
  \be  \bearl
  K^+_m := \ii\B m = \ii\! \sumzh r \normord{\b 1r \b 2{m-r}} \,, \qquad\
  K^-_m := \ii\C m = \ii\! \sumzh r \normord{\c 1r \c 2{m-r}} \,, \\{}\\[-.8em]
  K^0_m := \sqrt2\,\je m = \sumzh r (\normord{\b 1r \c 1{m-r}}
  + \normord{\b 2r \c 2{m-r}}) \,.\eear  \Labl19
(In a different context, this has also been observed in \cite{beps}.)

Next, let us study the observables which stay fixed under \dn, i.e.\ those
which are the observables of the orbifold and \sonh\ theories.
According to the results of \srf Q, we need to implement the
additional \zz\ transformation which operates on the fermion modes
as the exchange $\bogzz$ \Erf zz. We first note that $\bogzz$ acts on the
\glnh\ current modes \erf J as
  \be  \bogzz(\J ijm) = -\J jim  \,,  \ee
so that the invariant combinations are
  \be  \jo ijm := \J ijm - \J jim  \Labl jo
for $m\iN\zet$ and $i,j\iN\{\oneton\}$ with $i\,{<}\,j$. It follows from
the commutation relations \Erf53 that these modes span a level two \sonh\
affine \kma. This is also easily understood by realizing that the real
and imaginary parts $\u ir$ and $\v ir$ of the fermion modes, defined by
  \be  \b ir := \Frac1{\sqrt2}\,(\u ir+\ii\v ir) \,,  \qquad
       \c ir := \Frac1{\sqrt2}\,(\u ir-\ii\v ir) \,,  \labl{realc}
which satisfy $\bogzz(\u ir)\eq\u ir$ and $\bogzz(\v ir)\eq-\v ir$,
constitute the Fourier modes of a two sets of real free fermions. Each of these
realizes a real \cara, and they mutually anticommute. The combinations \Erf jo
are expressed through these fermion modes by
  \be  \bearll  \jo ijm \!\!&
  = \half \sumzh r(\normord{\u ir\u j{m-r}} - \normord{\u jr\u i{m-r}}
    +\normord{\v ir\v j{m-r}} - \normord{\v jr\v i{m-r}}) \\{}\\[-.8em]&
  = \sumzh r(\normord{\u ir\u j{m-r}}+\normord{\v ir\v j{m-r}})
  \,.  \eear \ee
It then follows immediately (compare e.g.\ \ci) that we are indeed dealing with
a level two \sonh\ \aff.

Now we use again the knowledge that the basic
observables of a \wzwt\ are given by the currents. {}From the inclusion
\erf{sub} we therefore conclude that the observables
\orbn\ of the orbifold theory are contained in the commutant of the current
\alg\ in the \alg\ $\car\odihed$. As this is a proper inclusion, we cannot
completely determine the orbifold observables this way. Nevertheless we can
make a few observations which can be compared to the literature.
First, the \ueh\ current modes \Erf je, which
in terms of the real fermions read $\je m\eq\ii(2N)^{-1/2}\Sumzh r\sumi\v ir
\u i{m-r}$, transform as $\bogzz(\je m)=-\je m$ and hence definitely
do not belong to \orbn. Also, among the
linear combinations of the \ce\ observables \Erf BC, only
$\B m+\C m$ are $\bogzz$-invariant; comparison with the results of \cite{dvvv}
shows that they belong indeed to the basic unbounded observables of the orbifold
theory, and hence the corresponding bounded functions are elements of $\orbn$.
However, there are still further basic
observables. These include in particular the \emt, whose modes can be written as
  \be   L_m^{{\sss({\sf Orb})}} = L_m^\caR - L_m^{\sss(\SO)} \ee
(analogously to $L_m\Inda\eq L_m^\caR\mi L_m\Inds$), where $L_m^{\sss(\SO)}$ is
obtained from the \sonh currents \Erf jo by the Sugawara formula. In terms of 
the fermion modes, this is a normal-ordered product which contains the fermion 
modes only in the quadratic form $\u ir\u is$ and $\v jr\v js$. According to
\cite{dvvv} there is one other
primary conformal field which is a basic observable, namely the combination
$\normord{\Je^4}-2\normord{\Je\partial^2\Je}+3/2\normord{(\partial\Je)^2}$.

In the special case $N\eq1$ the current \alg\ \sonh\ degenerates to \ueh,
with modes $\jo12m\eq\Sumzh r(\normord{\u1r\u2{m-r}}+\normord{\v1r\v2{m-r}})$,
or in terms of the \suzh\ modes \Erf18, $\jo12m\eq J^+_m-J^-_m$.
Similarly, the orbifold observables $\B m+\C m$ become
$\B m+\C m\eq-\ii(K^+_m+K^-_m)\eq\Sumzh r$\linebreak[0]$(\normord{\u1r\u2{m-r}}
-\normord{\v1r\v2{m-r}})$, which again generate a \ueh\ current \alg.
In fact, in this case we are dealing with
the tensor product of two \ce\ circle theories, each correponding to the value
$N\inda\eq4$ of the integer that labels the circle theories.

Finally we note that by construction the real fermions $\u i{}$ and $\v i{}$
are of Neveu\hy Schwarz type. In terms of the orbifold and \sonh\ theories,
this corresponds to the fact that we are dealing with untwisted sectors only.
In order to investigate the twisted sectors as well, one would have to include
also real fermions of Ramond type. Since for generic $N$ the twisted sectors
have non-integral statistical dimension, they are not covered by the
conventional DHR formalism, and hence are definitely
beyond the scope of our present paper.
On the other hand, when $N$ is a square number, then the statistical dimension
of the twisted sectors is integral, and correspondingly an interpretation in
terms of a DHR gauge group might again exist. At present we do not know of
such an interpretation. But it is easy to see that such a gauge group would
have to be an extension of \dn\ by \zz. Moreover, it is likely that this
extension should be central, in such a way that the twisted sectors
can be interpreted as projective \rep s of the factor group \dn.
(For $N\eq1$, this possibility is realized rather trivially as
the extension from $\dne\equiv\zet_4$ to $\zet_8$.)

\appendix

\sect{The finite groups \dn\ and \Dihed}

For any positive integer $N$, the {\em generalized quaternion group\/}
\dn\ is by definition the discrete group that
is generated freely by elements $x$ and $y$ modulo the relations\,%
\futnote{The first of these relations is not independent. We keep it to
demonstrate the similarity with \Erf3b below.}
  \be x^{2N}=1 \,, \qquad xyx=y \,, \qquad y^2=x^N \,. \Labl3a
This is a finite group of order $|\dn|=4N$. It has four one-dimensional
representations $\pi_\rmo,\pi_\rmv,\pi_\rms,\pi_\rmc$, as well as $N\mi1$
two-dimensional representations $\pi_m$ with representing matrices \cite{CUre}
  \be  \pi_m (x) = \left( \begin{array}{cc} \eE^{\ii\pi m/N} & 0 \\
  0 & \eE^{-\ii\pi m/N} \eear \right) , \qquad\quad
  \pi_m (y) = \left( \begin{array}{cc} 0 & (-1)^m \\ 1 & 0 \eear \right) . \ee
The conjugacy classes and characters of \dn\ are displayed in table \ref{T1}.
\begin{table}[bpth]\caption{Character table for \dn} \label{T1}
\begin{center}
  \begin{tabular}{|c||c|c|c|c|c|} \hline &&&&&\\[-.9em]
  {\footnotesize class} & $\,\{1\}\,$
    & $\{ x^k,x^{-k}\} \hsp{2.66}\atop (k\in\{1,2,...,N-1\})$ & $\{x^N\}$
    & $\!{\{ yx^{2l} \,|\hsp{2.62}\atop l=0,1,...,N-1 \}}\!$
    & $\!{\{yx^{2l+1}\,|\hsp{1.93}\atop l=0,1,...,N-1 \}}\!$
  \\[-.9em]&&&&&\\ \hline\hline &&&&&\\[-.9em]
  $\chi_\rmo$& $1$ & $1$ & $1$ & $\;1$  &  $\;1$ \\ &&&&&\\[-.9em]
  $\chi_\rmv$& $1$ & $1$ & $1$ & $-1\ $ & $-1\ $ \\ &&&&&\\[-.9em]
  $\chi_\rms$& $1$ &$(-1)^k$ & $(-1)^N$ & $\;\ii^N$ &$-\ii^N\ $
\\&&&&&\\[-.9em]
  $\chi_\rmc$& $1$ &$(-1)^k$ & $(-1)^N$ & $-\ii^N\ $ &$\;\ii^N$ \\&&&&&\\[-.9em]
  $\chi_m^{}$& $2$ & $2\cos(\pi mk/N)$& $2\,(-1)^{m} $ & $0$ & $0$
  \\[-.8em]&&&&&\\ \hline \multicolumn4c {} \\[.05em] \end{tabular}
\end{center} \end{table}
{}From the character table
it follows in particular that the subring of the \rep\ ring that is
furnished by the \onedim\ \rep s is the group ring of $\zz\Times\zz$ when
$N$ is even, and the group ring of $\zet_4$ when $N$ is odd.

The cyclic group \zn\ generated by $x$ is a normal subgroup of \dn;
\dn\ is a non-split extension of this normal subgroup by \zz.
It is illustrative to compare \dn\ to the {\em dihedral group\/} \Dihed\ which
is a {\em split\/} extension of its normal subgroup \zn\ by \zz\ and hence a
semi-direct product.
\Dihed\ is by definition generated by elements $\x,\,\y$ subject to the
relations
  \be  \x^{2N}=1 \,, \qquad \x\y\x=\y \,, \qquad \y^2=1 \,. \Labl3b
We have $|\dihed|=4N$, and there are four one-dimensional
representations $\pit_\rmo,\pit_\rmv,\pit_\rms,\pit_\rmc$, and $N-1$
two-dimensional representations $\pit_m$ with matrices
  \be  \pit_m (\x) = \left( \begin{array}{cc} \eE^{\ii\pi m/N} & 0 \\
  0 & \eE^{-\ii\pi m/N} \eear \right) , \qquad\quad
  \pit_m (\y) = \left( \begin{array}{cc} 0 & 1 \\ 1 & 0 \eear \right) . \ee
The conjugacy classes and characters of \Dihed\ are given in table \ref{T2}.
\begin{table}[bpth]\caption{Character table for $\dihed$} \label{T2}
\begin{center}
  \begin{tabular}{|c||c|c|c|c|c|} \hline &&&&&\\[-.9em]
  {\footnotesize class} & $\,\{1\}\,$
    & $\{ \x^k,\x^{-k}\} \hsp{2.66}\atop (k\in\{1,2,...,N-1\})$ & $\{\x^N\}$
    & $\!{\{ \y\x^{2l} \,|\hsp{2.62}\atop l=0,1,...,N-1 \}}\!$
    & $\!{\{\y\x^{2l+1}\,|\hsp{1.93}\atop l=0,1,...,N-1 \}}\!$
  \\[-.9em]&&&&&\\ \hline\hline &&&&&\\[-1.1em]
  $\chi_\rmo$ & $1$ & $1$ & $1$ & $\;1$ & $\;1$ \\ &&&&&\\[-.9em]
  $\chi_\rmv$ & $1$ & $1$ & $1$ & $-1\;$ & $-1\;$ \\ &&&&&\\[-.9em]
  $\chi_\rms$ & $1$ & $(-1)^k$ & $(-1)^N$ & $-1\;$ & $\;1$ \\ &&&&&\\[-.9em]
  $\chi_\rmc$ & $1$ & $(-1)^k$ & $(-1)^N$ & $\;1$ & $-1\;$ \\ &&&&&\\[-.9em]
  $\chi_m^{}$ & $2$ & $2\cos(\pi mk/N)$ & $2(-1)^{m}$ & $\;0$ & $\;0$
  \\[-.8em]&&&&&\\ \hline \multicolumn4c {} \\[.5em] \end{tabular}
\end{center} \end{table}
It can be checked that the \rep\ rings of both \dn\ and \Dihed\
are simply reducible.

Note that for even $N$ the groups \dn\ and \Dihed\ possess identical character
tables, and hence in particular identical \rep\ rings.
Nevertheless they are not isomorphic; e.g.\ in \dn\ there is only a single
element, namely $y^2$, of order two, while in \Dihed\ there are many.

\sect{Characters}

In \cite{bofu} we considered two species $\u i{}$ and $\v i{}$ of real free
fermions. An action of the group $\Ozw$ (which can be interpreted as a
DHR gauge group of the free fermion theory) was defined by
 \be  \gamma_t (\u ir) = \cos (t)\, \u ir - \sin (t)\, \v ir \,,\qquad
 \gamma_t (\v ir) = \sin (t)\, \u ir + \cos (t)\, \v ir \,,  \ee
where $t\iN\reals$, and
 \be  \eeta (\u ir) = \u ir \,, \qquad \eeta (\v ir) = - \v ir \,.  \ee
In terms of the complex fermions $\b ir$ and $\c ir$
(see formula \erf{realc}) this reads
 \be  \gamma_t (\b ir) = \eE^{\ii t}\, \b ir \,,\quad
 \gamma_t (\c ir) = \eE^{-\ii t}\, \c ir \,, \qquad
 \eeta (\b ir) = \c ir \,, \quad \eeta (\c ir) = \b ir \,.  \ee
It is obvious that by restricting this action to the naturally embedded
discrete subgroup $\dihed\subseT\Ozw$ we just recover the automorphisms
\erf{az} and \erf{zz}: $\bogzn1\eq\gamma_{\pi/N},\,\bogzz\eq\eeta$.

As in \ci\ we will use the labelling
  \be \Ozwh = \{0,J\} \cup \natnum  \ee
for the group dual of $\Ozw$.  In \cite{bofu} we computed
the characters of the sectors of $\car^\Ozw$; they read
  \be \bearl
  \chio 0 = \dstyle\frac{\sN 0}{2(\vi(q))^{2N}} +
  \frac{(\vi(q))^{2N}}{2(\vi(q^2))^{2N}} \,, \qquad
  \chio J = \frac{\sN 0}{2(\vi(q))^{2N}} -
  \frac{(\vi(q))^{2N}}{2(\vi(q^2))^{2N}} \,, \\{}\\[-.9em]
  \chio m = \dstyle\frac{\sN m}{(\vi(q))^{2N}} \quad{\rm for}\ m\iN\natnum\,,
  \eear \ee
with $\vi$ and $\SN m$ as defined in \Erf vi and \Erf sN. Now upon induction
from \Dihed\ to $\Ozw$ the irreducible \Dihed-\rep s split
into irreducible $\Ozw$-\rep s as
  \be \bearl
  \ind \Ozw \dihed \pidio = \pio_0 \oplus \bigoplus_{n=1}^\infty
  \pio_{2nN} \,,\qquad
  \ind \Ozw \dihed \pidiv = \pio_J \oplus \bigoplus_{n=1}^\infty
  \pio_{2nN} \,,\\[.6em]
  \ind \Ozw \dihed \pidis = \ind \Ozw \dihed \pidic = \bigoplus_{n=1}^\infty
  \pio_{(2n-1)N} \,,\\[.6em]
  \ind \Ozw \dihed {\pidi_\alpha} =  \bigoplus_{n=1}^\infty
  \left( \pio_{2(n-1)N+\alpha} \oplus \pio_{2nN-\alpha} \right)
  \quad{\rm for}\ \alpha\iN\{\onetond\}\,. \eear \ee
As a consequence, we have $\chido = \chio 0 +\sum_{n=1}^\infty \chio {2nN}$
etc., and hence we find the expressions
  \be \bearl
  \chido (q) = \psif N0\, \Frac{\sN 0}{2 (\vi(q))^{2N}} +
  \Frac{(\vi(q))^{2N}}{2 (\vi(q^2))^{2N}} \,,\\[.8em]
  \chidv (q) = \psif N0\, \Frac{\sN 0}{2 (\vi(q))^{2N}} -
  \Frac{(\vi(q))^{2N}}{2 (\vi(q^2))^{2N}} \,,\\[.8em]
  \chids (q) = \chidc (q) = \Frac12\, q^{-N/4} \psif NN\,
  \Frac{\sN N}{(\vi(q))^{2N}} \,,\\[.8em]
  \chid \alpha (q) = q^{-\alpha^2/4N} \psif N\alpha\,
  \Frac{\sN \alpha}{(\vi(q))^{2N}}\,, \quad \alpha\iN\{\onetond\}\,, \eear \ee
with $\Psif Na$ as defined in \erf{psif},
for the characters of the $\car^\dihed$ sectors. By comparison
with the characters of the sectors of $\orbn$ and $\wzwo$ (given e.g.\ in
\cite{bofu}), it is then easily checked that indeed the identities \erf{mess}
hold. Moreover, by restricting further the $\dihed$-action to the $\Zz$-action
of $\bogzz$ (corresponding to the element $\y$ of \Dihed), we obtain
  \be \bearl
  \ind \dihed \Zz {\pizzp}  = \pidio \oplus \pidic \oplus
  \bigoplus_{\alpha=1}^{N-1} \pidi_\alpha \,,\\[.8em]
  \ind \dihed \Zz {\pizzm}  = \pidiv \oplus \pidis \oplus
  \bigoplus_{\alpha=1}^{N-1} \pidi_\alpha \,, \eear \ee
where $\pizzp\equiv\id$ and $\pizzm$ denote the two irreducible
representations of $\Zz$. The identities \erf{evod} then immediately follow as
a consequence of \erf{mess}.

\bigskip\bigskip\bigskip{\small
\noindent{\bf Acknowledgement.}\\ It is a pleasure to thank K.-H.\ Rehren for
helpful comments on the manuscript.}
\newpage


  \newcommand\wb       {\,\linebreak[0]} \def\wB {$\,$\wb}
  \newcommand\Bi[1]    {\bibitem{#1}}
  \newcommand\BOOK[4]  {{\em #1\/} ({#2}, {#3} {#4})}
  \renewcommand\J[5]   {\ {\sl #5}, {#1} {#2} ({#3}) {#4} }
  \newcommand\JJ[5]    {{\sl #5}  {#1} {#2} ({#3}) {#4}}
 \def\jf    {J.\ Fuchs}
 \def\coma  {Con\-temp.\wb Math.}
 \def\comp  {Com\-mun.\wb Math.\wb Phys.}
 \def\ijmp  {Int.\wb J.\wb Mod.\wb Phys.\ A}
 \def\jomp  {J.\wb Math.\wb Phys.}
 \def\npbp  {Nucl.\wb Phys.\ B (Proc.\wb Suppl.)}
 \def\nupb  {Nucl.\wb Phys.\ B}
 \def\rims  {Publ.\wB RIMS}
 \def\rmap  {Rev.\wb Math.\wb Phys.}
 \def\A       {Algebra}
 \def\Be     {{Berlin}}
 \def\Con     {Conformal\ }
 \def\cua     {current algebra}
 \def\fts     {field theories}
 \def\furu    {fusion rule}
 \def\NY     {{New York}}
 \def\oa      {operator algebra}
 \def\q       {quantum\ }
 \def\Q       {Quantum\ }
 \def\con     {conformal\ }
 \def\qg      {quantum group}
 \def\modinv  {modular invarian}
 \def\inv     {invariance}
 \def\Rep     {Representation}
 \def\SV     {{Sprin\-ger Verlag}}
 \def\sym     {symmetry}
 \def\syms    {sym\-me\-tries}
 \def\trfo    {transformation}
 \def\wzw     {WZW\ }
 \def\stc     {statistic}
 \def\eq      {equa\-tion}
 \def\dimn    {dimension}
 \def\WZ      {Wess\hy Zu\-mino }

\version\versionno
\end{document}